# Spin wave imaging in atomically designed nanomagnets

A. Spinelli[1], B. Bryant[1], F. Delgado[2], J. Fernández-Rossier[2], and A. F. Otte[1*]

[1]Department of Quantum Nanoscience, Kavli Institute of Nanoscience, Delft University of Technology, Lorentzweg 1, 2628 CJ Delft, The Netherlands

[2]International Iberian Nanotechnology Laboratory (INL), Avenida Mestre José Veiga, 4715-310 Braga, Portugal

* a.f.otte@tudelft.nl

**The spin dynamics of all ferromagnetic materials are governed by two types of collective excitations: spin waves and domain walls. The fundamental processes underlying these collective modes, such as exchange interactions and magnetic anisotropy, all originate at the atomic scale; yet, conventional probing techniques, based on neutron [1] and photon scattering [2], provide high resolution in reciprocal space, and thereby poor spatial resolution. Here we present direct imaging of spin waves in individual chains of ferromagnetically coupled $S = 2$ Fe atoms, assembled one by one on a $Cu_2N$ surface using a scanning tunnelling microscope. We are able to map the spin dynamics of these designer nanomagnets with atomic resolution, in two complementary ways. First, atom to atom variations of the amplitude of the quantized spin wave excitations, predicted by theory, are probed using inelastic electron tunnelling spectroscopy. Second, we observe slow stochastic switching between two opposite magnetisation states [3,4], whose rate varies strongly depending on the location of the tip along the chain. Our observations, combined with model calculations, reveal that switches of the chain are initiated by a spin wave excited state which has its antinodes at the edges of the chain, followed by a domain wall shifting through the chain from one end to the other. This approach opens the way towards atomic scale imaging of other types of spin excitations, such as spinons and fractional end-states [5,6], in engineered spin chains.**

When a magnetic domain reverses its magnetisation, either due to thermal fluctuations or induced by an electric current [7,8], collective spin excitations inside the magnet follow each other at very fast timescales, until a stable magnetic situation has been regained. Insight into which collective states are populated at what stage of the reversal process contributes to our understanding of ultrafast spin dynamics, and is essential for the development of current-switched magnetic storage devices [9].

Analogous to acoustic phonon modes in a solid-state material, the lowest energy excitations of a finite one-dimensional chain of coupled spins are long-wavelength spin waves, whose energies are quantised due to confinement inside the chain. Quantum-mechanically, a spin wave can be seen as a magnon: a single $\Delta m = 1$ excitation that can freely propagate. In a higher order spin wave, the magnon is not completely delocalised along the chain, but is instead confined to the antinodes of the wave. For even higher energies yet more complex states exist, such as multiple magnon states or domain wall states, in which entire sections of the chain have flipped their magnetisation. To date, no atomic resolution imaging of spin waves has been reported.

Recent advancements in the field of low temperature scanning tunnelling microscopy (STM) have made it possible to probe static spin textures on the atomic scale [10], as well as manipulate magnetic atoms into artificial assemblies such as spin chains [11] and spin-based logic gates [12]. The strength of the spin coupling between neighbouring spins inside these assemblies may be tuned by adjusting the relative positioning of the atoms [13–16]. For close-coupled atoms, the spin texture will be identical anywhere along the structure [4,11,17,18], whereas in the weak limit the coupling can be considered as only a small perturbation to the single spin case [19]. Our experiments allow us to access the intermediate case, in which the spins behave as a collective magnetic entity, but where each spin may still be addressed individually [3,16].

We use a 330 mK STM to assemble chains of Fe atoms on a $Cu_2N$ lattice along the [110] direction, in which configuration the exchange coupling $J = -0.7$ meV between neighbouring atomic spins $S = 2$ is ferromagnetic [15] (Fig. 1a). When the chains are imaged using a non-magnetic STM tip, the tunnel current is stable. However, when we scan a chain of five or more atoms with a spin-polarised tip (which is achieved by replacing the last atom on the tip with an Fe atom, and applying a small magnetic field in the [100] direction [20]), the tip and the chain form a magnetic tunnel junction and stochastic switching between two conductance states is observed (see Figs. 1b and c). For shorter chains, the switching becomes faster than the bandwidth of our measurement. Since we typically measure in constant current mode, switching between the conductance states, which we ascribe to transitions between the two ferromagnetic metastable states of the chain, is observed as changes in the tip height: a reduction of the tip height by 6 pm corresponds to a decrease in conductance of approximately 10%.

As expected for ferromagnetically coupled spins in a magnetic field, one of the two states is favoured over the other; the asymmetry in the occupancies of the two states (i.e., the magnetisation) is found to increase with the strength of the field (Figs. 1d, S1 and S2). The favoured high conductance state corresponds to ($m = +2,+2,+2,+2,+2,+2$) ≡ GS and the unfavoured low conductance state to ($m = -2,-2,-2,-2,-2,-2$) ≡ GS*, where $m = +2$ ($-2$) refers to a single spin being aligned parallel (anti-parallel) to the external magnetic field (Fig. 1a).

Interestingly, we find that the switching rate depends strongly on which atom in the chain the STM tip is positioned over (Figs. 1c, 2b and S1). Using a bias voltage that is sufficiently large to drive inelastic spin excitations, we measure a decrease in the switching rate of approximately an order of magnitude when moving the tip from the outer atoms to the centre of the chain. In this paper we show how this strong position dependence of the switching rate, which was not observed in previous experiments [3,4], provides a new magnetic imaging technique that we use to investigate the intrinsic collective spin dynamics of the chain.

In order to identify the excited states of the spin chain we performed inelastic electron tunnelling spectroscopy (IETS) [11,21]: Fig. 2a shows spectra taken on each atom of a six atom chain (see Fig. S1 for a five atom chain). Allowed spin excitations appear as peaks in the second voltage derivative of the current ($d^2I/dV^2$) at voltages corresponding to the excitation energies. We observe that all the atoms of the chain share the same lowest energy excitation, at bias voltage $V = 3.6$ mV, with comparable intensity in $d^2I/dV^2$. For higher energy excitations, we see that the peak intensity can change from atom to atom, and from peak to peak, in a manner that is consistent with spin waves. In this way, by mapping the nodal structure of the spin-wave states as a function of energy, we demonstrate a method for measuring spin wave dispersion.

Analysis of the IETS spectra through comparison to a Spin Hamiltonian describing the anisotropy of the spins in combination with nearest neighbour Heisenberg coupling [15,19,22,23] (Figs. 2a and S3, see Methods), confirms that the lowest energy excitation ($V = 3.6$ mV) from GS* corresponds to a fully delocalised spin wave state: an equal superposition of all six permutations of ($m = -1,-2,-2,-2,-2,-2$) (Fig. 3a). In the following we will refer to this state as SW1.

The next excited state follows very shortly after, at 4.0 mV. As can be seen from the IETS spectra, the step belonging to this excitation is visible most prominently on atoms 1 and 6, and to a lesser extent also on atoms 2 and 5. According to our analysis, this state, labelled SW2, is a spin wave with one node in the centre of the chain: ($m = -1,-2,-2,-2,-2,-2$) and ($m = -2,-2,-2,-2,-2,-1$) are represented more strongly than permutations where the magnon resides away from the edges. The third excited state, SW3 at 5.1 meV, has two nodes: on atoms 2 and 5. The observed variations of IETS intensity provide an atomic resolution image of the spin wave excitations, and can be understood in terms of the compositions of the excited states, obtained by means of exact numerical diagonalization, that can be found in Table S1.

Fig. 2c shows how the measured switching rate from GS* to GS changes as the bias voltage is varied. We see that the rate increases rapidly as excitations into states SW1 and SW2 become accessible [24]. In addition, we find that for voltages far below the excitation threshold, the spatial dependence of the switching rates is markedly suppressed (see also Fig. S1). We therefore conclude that switching processes are initiated by tunnelling electrons making excitations into either of these two spin wave states, and that the spatial dependence is a result of current driven processes: the intrinsic switching of the chain (through thermal processes, induced by substrate electrons [25], or as a result of quantum dissipative tunnelling of magnetisation) is primarily tip-position independent. It should be noted that our bias-dependent measurements were performed with the current held constant, so that the tip height increases with bias voltage.

In order to gain further insight into the switching dynamics, we compared the experimentally observed switching behaviour to calculations based on Pauli master equations [20,26,27] (details in Methods). In these calculations, the position of the tip is modelled by allowing transitions that involve electrons tunnelling into and out of the tip on only one of the six spins: on the other spins, only transitions due to exchange coupling with substrate electrons are allowed. As shown in Fig. 2b, the spatial dependence of the switching rates is qualitatively reproduced by the theory.

The calculated spatial dependence of the switching rate has its origin in the sub-picosecond timescale following an initial excitation. In Figs. 3b and c we show the calculated evolution of the state occupancies, when the system is prepared at time $t = 0$ in SW1, with the tip positioned over atoms 1 and 3. In either case, with almost 100% certainty, the spins are found to relax back from SW1 into GS* within a few picoseconds. However, we see that there is a small probability that an excitation to SW1 sets up a chain of events involving temporary occupation of SW2 followed by other excited states, resulting in a collective switch of the spin chain into GS. When the tip is over atom 3, this probability is approximately $5\times10^{-7}$, whereas for atom 1 it is $9\times10^{-6}$, consistent with our findings of atom dependent switching rates. Since the excitation path always passes through the edge-localised state SW2, we conclude that the observed spatial dependence of switching rate is due to the nodal structure of this state.

In addition, the calculated probabilities reveal that most other excited states involved in the switching process are of the domain wall type. A clear sequence can be seen starting from DW1, where only one edge spin has flipped to $m = +2$, via states DW2 and DW3, gradually moving the domain wall through the chain, to GS (Fig. 3a). It is interesting to note that, in view of the quantum mechanical nature of these states, even though the tip is making spin excitations on only one end of the chain, the domain wall is actually propagating through the chain in both directions simultaneously.

The measurements as a function of bias voltage presented in Fig. 2c, taken at 20 pA tunnel current, indicate that even for voltages well below the excitation threshold, the switching rate still depends somewhat on the position of the tip along the chain. Subsequent measurements taken at a smaller current (and, consequently, a larger tip height) reveal that this switching is partly induced by the proximity of the tip (Fig. 4a). At a current of 5 pA, the switching rate has decreased to 3 mHz and is independent of the tip position along the chain. Extrapolation of the current dependence suggests an intrinsic switching rate of 1 mHz. For voltages near the excitation threshold, the position dependence is not influenced by the current. The observed current dependent switching below threshold could be caused by a tip induced magnetic or electric field [28].

At bias voltages well above 4 mV, the switching rates become too fast for conventional DC measurements. Nevertheless, by employing a pulsed measurement scheme [3] it is possible to use position-dependent switching rates to map the spatial structure of higher energy spin excitations. Fig. 4b shows switching rates acquired using pulsed voltages up to 21 mV. The data reveal a different spatial variation due to the influence of higher-order spin wave states such as SW3. At 7 mV, the combined effect of SW2 and SW3 leads to a switching rate, which is highest for atoms 1 and 6 and lowest for 2 and 5. For even higher bias voltages, the spatial variation is much diminished, due to the compensatory influence of centre localised states such as the one observed at 9.2 mV in Fig. 2a. These results demonstrate an additional method for mapping spin wave dispersion.

In summary, we have demonstrated direct visualization of spin waves in an atomically assembled spin chain using two complementary techniques: inelastic electron tunnelling spectroscopy and atomic scale magnetometry. The combination of these two approaches provides extraordinarily precise information on the microscopic mechanisms that drive switching between the two metastable states of a ferromagnetic chain. This method represents a new level of precision for testing theories of magnetic reversal, with relevance not only to further research in spin excitations, but also to the process of writing information to magnetic domains.

## METHODS

**Experimental parameters.** Unless specified otherwise, measurements were carried out at 330 mK in ultra-high vacuum (< $2\times10^{-10}$ mbar), in a commercial STM system (Unisoku USM-1300S). Magnetic fields were applied in the plane of the surface, parallel to the primary anistropy axis of the Fe atoms. The $Cu_2N$ substrate was prepared in situ by $N_2$ sputtering of a Cu(100) crystal [29]. Fe atoms were evaporated onto the precooled $Cu_2N$ surface. STM tips were prepared by indenting commercial Pt-Ir STM tips into the Cu surface. Fe chains were assembled on the $Cu_2N$ surface using vertical atom manipulation [11,16].

For IETS measurements, differential conductance ($dI/dV$) spectra were recorded at a tunnel impedance of 11 MΩ with a lock-in amplifier, using an excitation voltage of 70 $\mu V_{RMS}$ at 928 Hz. $d^2I/dV^2$ spectra were obtained through numerical differentiation. Telegraph noise data were recorded at a constant current of 20 pA, unless otherwise specified. Switching rates were measured from exponential fits to histograms of the residence times in each state [3]. For measurements using pulsed bias (Fig. 4), a DC bias of 1 mV was applied at 20 pA; bias pulses of 6, 8, 10 and 20 mV were then added, with widths varying from 5 to 100 ns, at a repetition rate of 20 ms [3]. In order to measure switching rates at biases well below the threshold for inelastic excitations, we used a feedback mechanism as shown in Fig. S4.

**Spin Hamiltonian.** The Fe chains are modelled by a Heisenberg spin Hamiltonian of spins $S = 2$ with first neighbour interactions $J$, and single ion magnetic anisotropy with axial $D$ and transverse $E$ terms [11,23] (see Fig. S3). These parameters are determined by fitting experimental $dI/dV$ spectra [11,22]. From fitting the $N = 6$ chain at zero field we find $D = -1.29$ meV, $E = 0.31$ meV, $J = -0.73$ meV, very close to values reported in previous work [15,22].

**Master equation calculations.** The analysis of the dynamical evolution is done using a master equation description for the time evolution of the occupations of the energy levels of the spin chain [26]. The scattering rates depend on the spin matrix elements and on three dimensionless parameters: $(\rho_S J_K)^2$, which determines the strength of the exchange coupling to the surface, and $\rho_S \rho_T T_J^2$ and $\rho_S \rho_T T^2$, which settle the spin dependent and spin independent tunnelling amplitudes respectively. Here $\rho_S$ and $\rho_T$ are the surface and tip density of states at the Fermi energy, $J_K$ is the exchange with the surface, $T_J$ is the tunnelling exchange [23], and $T$ is the elastic tunnelling amplitude. The ratio between the zero-bias conductance and the inelastic contribution permits determining the ratio $T_J/T$ [20,26]. The parameter of tunnelling amplitude $\rho_S \rho_T T_J^2$ is obtained from the total current $I$ =20 pA while the switching time from GS* to GS at bias voltage $V$ = 4.2 mV and the tip positioned over atom 1 is used to extract $(\rho_S J_K)^2$. From the fitting we get $(\rho_S J_K)^2 = 3.8\times10^{-2}$, consistent with values previously found for Co atoms on $Cu_2N$ [30], $\rho_S \rho_T T_J^2 = 1.1\times10^{-5}$ and $T_J/T = 0.65$.

**Acknowledgements**
This work was supported by the Dutch funding organizations FOM and NWO (VIDI) and by the Kavli Foundation. F.D. and J.F.R. acknowledge support from the Ministry of Science and Education Spain (FIS2010-21883-C02-01) and from GV grant Prometeo (ACOMP/2010/070).

**Author contributions**
A.S. and B.B. performed the measurements and analysed the results. F.D. and J.F.R. performed the master equation calculations and provided theoretical support. A.F.O. conceived the experiment and supervised the work. All authors contributed to writing the manuscript and gave critical comments.

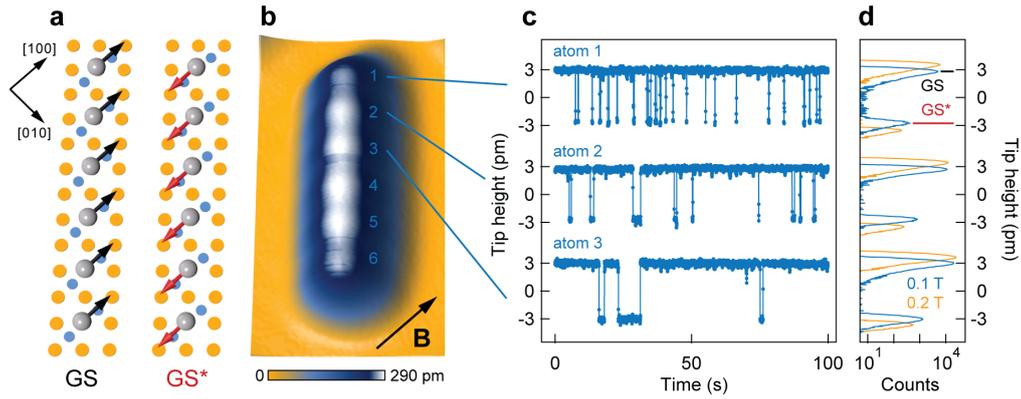

**Figure 1 | Magnetic switching of a ferromagnetic spin chain. a**, Atomic configuration and magnetic ground state orientations of a ferromagnetically coupled chain of six Fe atoms (gray) on $Cu_2N$ (Cu=yellow, N=blue). The magnetic easy axes of the Fe atoms are oriented along the [100] direction, giving rise to two collective metastable magnetic states GS (black) and GS* (red). The distance between neighbouring Fe atoms in the chain is 0.51 nm. **b**, STM topography taken with a spin-polarised tip (2.5 nm × 4.5 nm, 20 pA, 4.2 mV) in a magnetic field $B$ = 200 mT along the [100] direction. Switches between GS and GS* may be observed in the image. **c**, Constant current measurements (20 pA, 3.7 mV) of the tip height as a function of time for atoms 1, 2 and 3 in the chain at 100 mT. **d**, Tip height histograms, taken on the same three atoms, for field values of 100 mT and 200 mT (20 pA, 3.7 mV). Each histogram is based on a measurement time containing at least 80 switches in each direction.

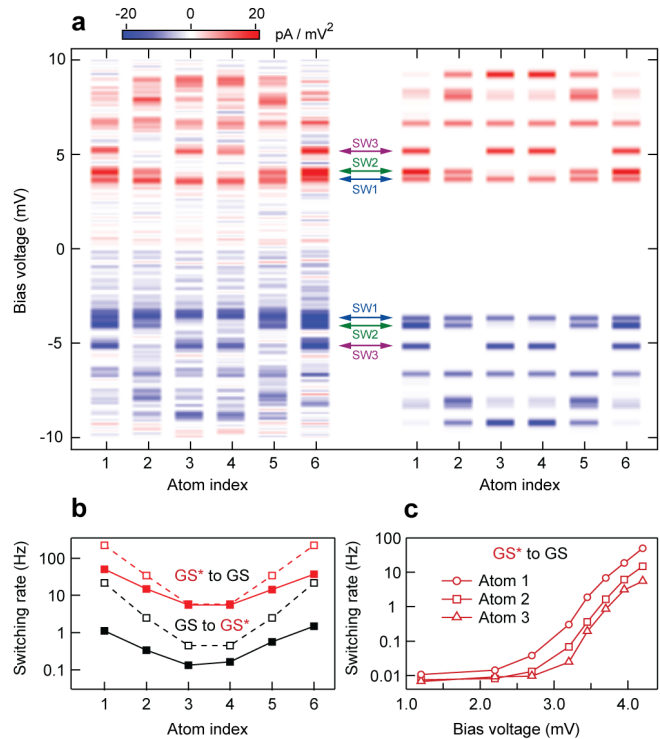

**Figure 2 | Observation of spin wave states. a**, Left: IETS spectra ($d^2I/dV^2$) taken on each atom of a ferromagnetically coupled Fe chain at $B$ = 200 mT. Right: corresponding simulated spectra. Spin excitations appear as bright red and blue bars. The first three spin waves are labelled. Corresponding $dI/dV$ spectra are shown in Fig. S3. **b**, Experimental (solid markers) and theoretical (open markers) telegraph noise switching rates as a function of position along the chain, at 4.2 mV. All switching rates are based on time traces containing at least 150 switches in each direction. **c**, Observed switching rate from GS* to GS as a function of applied bias voltage. Rates below 3.5 mV were measured using a feedback mechanism, see supplementary figure S4. All telegraph noise data were taken in constant current mode at 200 mT: measurement uncertainties are all smaller than the data symbols.

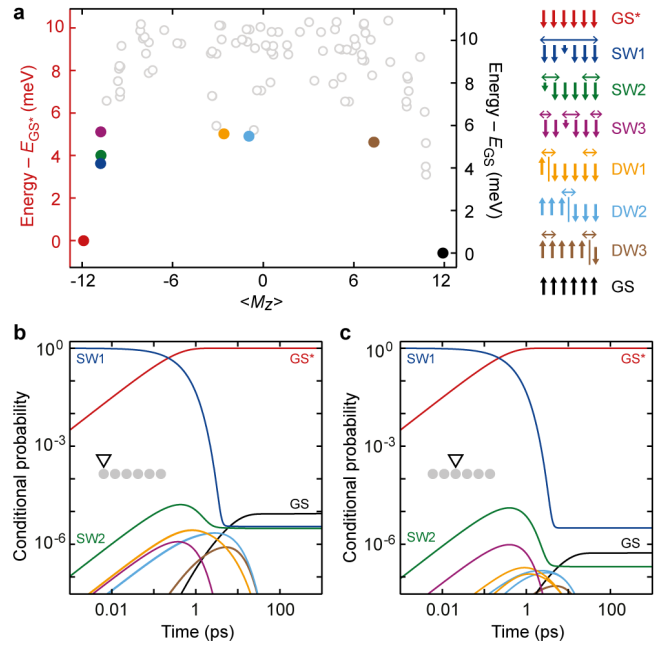

**Figure 3 | Time evolution of the switching process. a**, Energy vs. $<M_Z>$, the projection of the total magnetic moment in the direction of the field for the 100 lowest energy states at $B$ = 200 mT. States playing a key role in the switching process from GS* to GS are coloured and schematically represented in the legend. **b**, Calculated occupation probabilities of eigenstates following initialization in state SW1, with the tip positioned over atom 1, a bias voltage of 4.2 mV and a current of 20 pA. The colours refer the legend of **a**. For the time evolution beyond 1 ns, refer to Fig. S5. **c**, Same as **b**, for tip over atom 3.

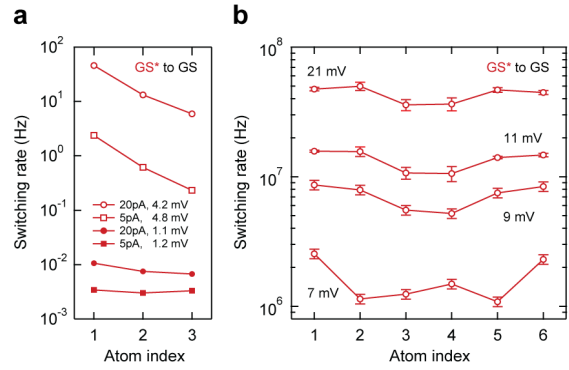

**Figure 4 | Current and bias dependences of the switching rates. a**, Dependence of switching rate on the tip position along a six-atom Fe chain, both above and below the bias voltage threshold for inelastic excitations of 3.6 mV, for different tunnel currents. Sub-threshold switching rates were measured using a feedback mechanism (see Fig. S4), so only the GS* to GS rate is recorded. Because the sub-threshold measurements at low current are extremely slow, only the first three atoms were measured. **b**, Dependence of the switching rate on tip position for bias voltages above 5 mV, recorded using a pulsed bias at 20 pA. Since the tip height was kept constant during the pulsed measurements, these switching rates are not to be compared directly to those taken in the constant current DC mode. All data taken at $B$ = 200 mT.

# Spin wave imaging in atomically designed nanomagnets

A. Spinelli[1], B. Bryant[1], F. Delgado[2], J. Fernández-Rossier[2], and A. F. Otte[1]

[1]*Department of Quantum Nanoscience, Kavli Institute of Nanoscience, Delft University of Technology, Lorentzweg 1, 2628 CJ Delft, The Netherlands*

[2]*International Iberian Nanotechnology Laboratory (INL), Avenida Mestre José Veiga, 4715-310 Braga, Portugal*

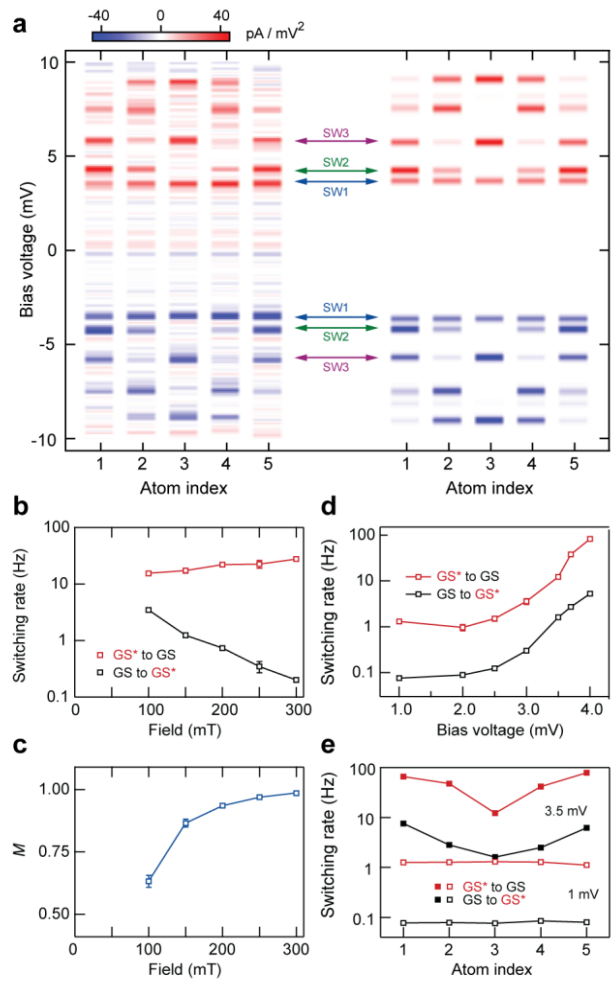

**Figure S1 | Five atom ferromagnetic Fe chain**. **a**, Left: IETS spectra ($d^2I/dV^2$) taken on each atom of a five atom Fe chain at zero external magnetic field and 330 mK. Right: corresponding simulated spectra. Spin excitations appear as bright red and blue bars. The energies of states SW2 and SW3 have increased slightly compared to the six atom chain, due to tighter confinement. **b**, Telegraph noise switching rates in both directions as a function of the external magnetic field, measured on the central atom of the chain, at 3.5 mV and a constant current of 20 pA. **c**, Nonequilibrium magnetisation $M = (\tau_+ - \tau_-)/(\tau_+ + \tau_-)$, where $\tau_+$ and $\tau_-$ are the reciprocal of the switching rates from GS to GS* and GS* to GS respectively [1], corresponding to the data of panel **b**, as a function of magnetic field. **d**, Bias voltage dependence of both switching rates at 150 mT, measured on atom 3 and at a constant current of 20 pA. **e**, Switching rates in either direction vs. atom index, above (3.5 mV) and below (1 mV) the threshold for inelastic excitations. No spatial dependence is observed at low bias voltage. All the data shown in panels **b**-**e** were measured at a temperature of 1.5 K: error bars are not shown when smaller than the markers.

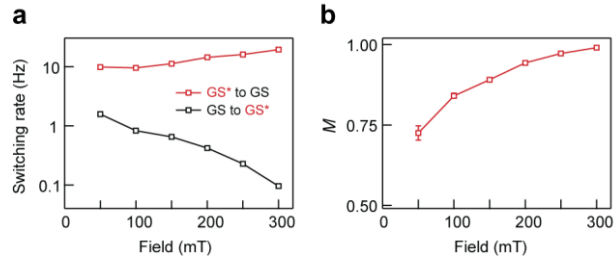

**Figure S2 | Six-atom Fe chain: magnetic field dependence. a**, Telegraph noise switching rates as a function of the external magnetic field, measured on one of the outer atoms of the chain, at 3.95 mV and a constant current of 20 pA, at 330 mK. The opposite behaviour of the two rates should be noted, in agreement with an increase of the magnetisation with magnetic field. **b**, Nonequilibrium magnetisation $M$, corresponding to the data of panel **a**, as a function of magnetic field. At 300 mT, the structure is completely magnetised in the direction of the field. Error bars are not shown when smaller than the markers.

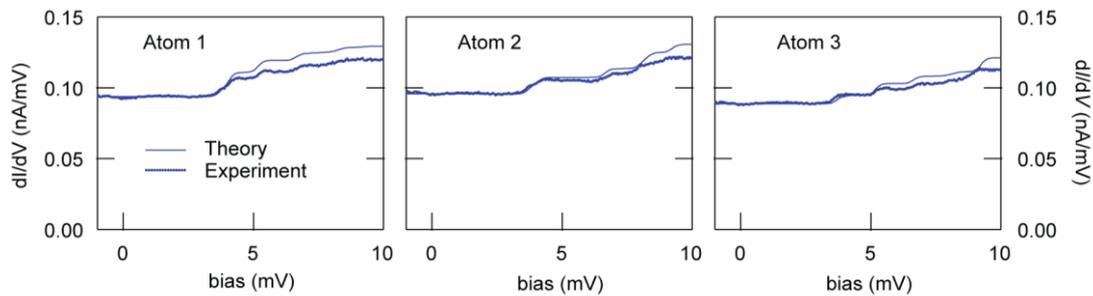

**Figure S3 | Six-atom Fe chain: d$I$/d$V$ spectra.** Data from Fig. 2, presented as d$I$/d$V$. Only the first three atoms are shown. Points: experimental data, 330 mK, $B$ = 200 mT. Lines: calculated spectra based on $D = -1.29$ meV, $E = 0.31$ meV, $J = -0.73$ meV.

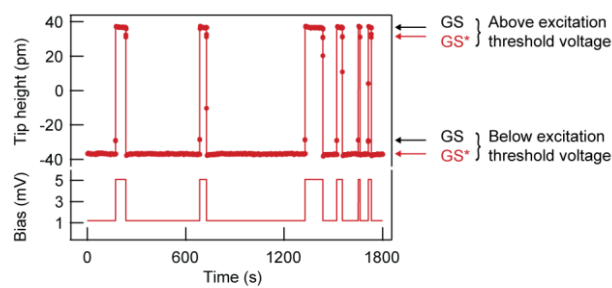

**Figure S4 | Method to measure slow switching rates.** A feedback mechanism is used to measure the intrinsic GS* to GS switching rate. In order to speed up the measurement, the (slow) switching rate from GS to GS* is measured at a bias voltage above the inelastic excitation threshold, and the (faster) switching rate from GS* to GS is measured below the threshold voltage. As soon as a switch from GS to GS* in the high bias voltage state is recorded, the bias voltage is reduced to the lower value, with the system ending up in GS*, at the low bias voltage. When a switch to GS is detected, the bias voltage is restored to the high value. This procedure was repeated until at least 150 switches had been recorded. In this way we could extract the intrinsic (low-bias) switching rate from GS* to GS. This example trace was recorded at a current of 5 pA and a field of 200 mT, at 330 mK, on the second atom of a six-atom Fe structure: biases of 5.1 mV and 1.2 mV were used. This procedure was used to measure all switching rates below 3.5 mV in Figs. 2c and 4a.

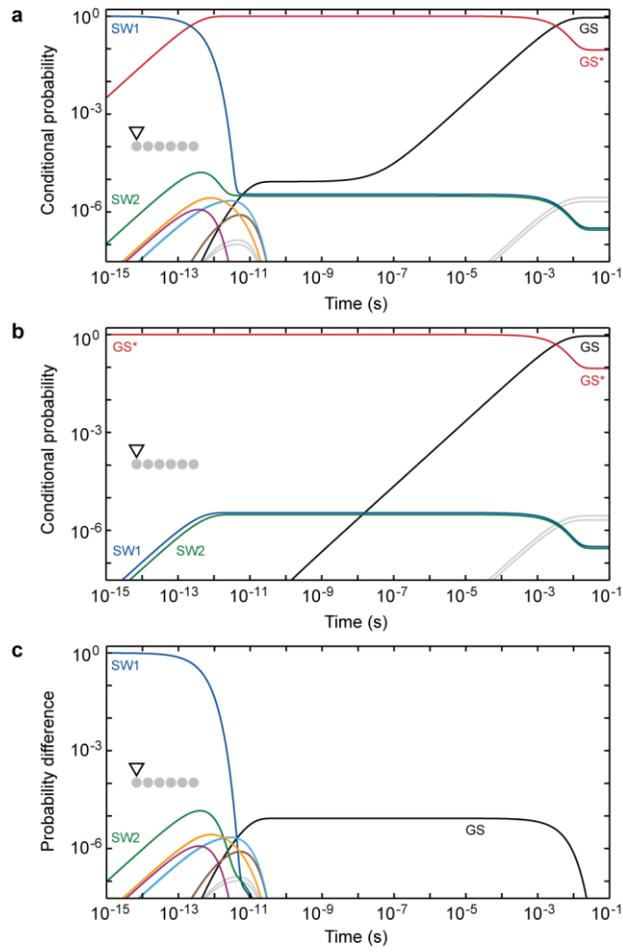

**Figure S5 | Long-term time evolution of the switching process.** **a,** Same as Fig. 3b, but showing a timescale up to 100 ms. **b,** same as **a**, but with initialization in GS* instead of SW1. **c**, Difference between **a** and **b** for all states. For times longer than 100 ps, the two initializations lead to identical time evolutions, except for the plateau of state GS at 9×10$^{-6}$, representing the additional probability of ending up in state GS as a result of a single excitation in SW1.

| State | Energy – $E_{GS^*}$ (meV) | +2+2+2+2+2+2 | –2–2–2–2–2–2 |
|---|---|---|---|
| GS | –0.58 | 0.95 | |
| GS* | 0 | | 0.95 |

| State | Energy – $E_{GS^*}$ (meV) | –1–2–2–2–2–2 / –2–2–2–2–2–1 | –2–1–2–2–2–2 / –2–2–2–2–1–2 | –2–2–1–2–2–2 / –2–2–2–1–2–2 |
|---|---|---|---|---|
| SW1 | 3.63 | 0.16 / 0.16 | 0.15 / 0.15 | 0.14 / 0.14 |
| SW2 | 4.01 | 0.29 / 0.29 | 0.14 / 0.14 | |
| SW3 | 5.12 | 0.21 / 0.21 | | 0.24 / 0.24 |

| State | Energy – $E_{GS^*}$ (meV) | +2–2–2–2–2–2 / –2–2–2–2–2+2 | +2+2–2–2–2–2 / –2–2–2–2+2+2 | +2+2+2–2–2–2 / –2–2–2+2+2+2 | +2+2+2+2–2–2 / –2–2+2+2+2+2 | +2+2+2+2+2–2 / –2+2+2+2+2+2 |
|---|---|---|---|---|---|---|
| DW1 | 5.03 | 0.17 / 0.17 | | | 0.12 / 0.12 | |
| DW2 | 4.91 | 0.06 / 0.06 | 0.10 / 0.10 | 0.15 / 0.15 | 0.06 / 0.06 | |
| DW3 | 4.63 | | | | 0.06 / 0.06 | 0.29 / 0.29 |

**Table S1 | State compositions.** Probability distributions (wave function squared) and corresponding energies of the eigenstates labelled in Fig. 3. Only components larger than 0.05 are shown.